\documentstyle[12pt]{article}
\def\setb@se#1{\baselineskip=#1 \normalbaselineskip=#1}
\setb@se{14pt}
\def\dal{\hbox{\hskip 0.5mm\hbox{\vrule width2.3mm height0.2mm
\vbox{\hrule width0.3mm height2.6mm}\hskip
-2.6mm \vbox{\hbox{\vrule width2.6mm height0.1mm}
\vskip -0.1mm\hrule width0.1mm height2.6mm}}\hskip 0.5mm}}
\newcommand{\be}{\begin{equation}}
\newcommand{\ee}{\end{equation}}
\newcommand{\bea}{\begin{eqnarray}}
\newcommand{\eea}{\end{eqnarray}}
\newcommand{\bdm}{\begin{displaymath}}
\newcommand{\edm}{\end{displaymath}}
\newcommand{\benu}{\begin{enumerate}}

\newcommand{\eenu}{\end{enumerate}}
\newcommand{\daltil}{\tilde{\dal}}

\newcommand{\nabtil}{\tilde{\nabla}}

\newcommand{\gtil}{\tilde{g}}
\newcommand{\ghat}{\hat{g}}

\newcommand{\gtens}{\mbox{\boldmath $g$}}

\newcommand{\ghattens}{\hat{\gtens}}
\newcommand{\gtiltens}{\tilde{\gtens}}

\newcommand{\Llow}{\Lambda_{\mbox{\scriptsize crit}}}
\newcommand{\Lmid}{\Lambda_{\star}}
\newcommand{\Lhig}{\Lambda_{\mbox{\scriptsize reg}}}

\input epsf

\begin{document}

\begin{titlepage}
\begin{flushright}
hep-th/9605089\\
ZU-TH 13/96\\
May 1996
\end{flushright}
 
\vfill

\begin{center}
{\huge Cosmological Analogues of the 

\vspace{2 mm}

Bartnik--McKinnon Solutions}

\vfill

{\bf M.S. Volkov, N. Straumann, 
G. Lavrelashvili\footnote{On  leave of absence from 
Tbilisi Mathematical
Institute, 380093 Tbilisi, Georgia},

M. Heusler and O. Brodbeck}

\vfill

{\em Institut f\"ur Theoretische Physik, Universit\"at Z\"urich}\\
{\em Winterthurerstrasse 190, CH--8057 Z\"urich, Switzerland}\\

\vfill

{\Large\bf Abstract}

\vfill

\end{center}

\noindent
We present a numerical classification of the spherically symmetric,
static solutions to the Einstein--Yang--Mills equations with
cosmological constant $\Lambda$.
We find three qualitatively different classes of configurations,
where the solutions in each class are characterized by the value
of $\Lambda$ and the number of nodes, $n$, of the Yang--Mills
amplitude.

For sufficiently small, positive values of the cosmological constant, 
$\Lambda < \Llow(n)$, the solutions generalize the Bartnik--McKinnon
solitons, which are now surrounded by a cosmological horizon and
approach the deSitter geometry in the asymptotic region.
For a discrete set of values $\Lambda_{\rm reg}(n) > 
\Lambda_{\rm crit}(n)$, the solutions are topologically $3$--spheres, the
ground state $(n=1)$ being the Einstein Universe. 
In the intermediate region, that is for $\Llow(n) < \Lambda < \Lhig(n)$,
there exists a discrete family of global solutions with
horizon and ``finite size''.

\vfill 

\end{titlepage}

%
\section{Introduction}

The interplay of gravity and non--linear field theoretical matter models 
leads to a wealth of new and surprising phenomena. 
In particular, there has been an increasing interest in both the
structure and the stability of black hole solutions ``with hair''.
(See, e.g., \cite{BHS96} and \cite{mavr} for some key references.)
Moreover, self--gravitating field theories have also become very popular in 
cosmology in connection with various inflationary scenarios, the 
formation of topological defects in cosmological phase transitions, etc.
 
In this paper we present and discuss some new solutions with various 
global properties of the Einstein--Yang--Mills (EYM) system with
cosmological constant $\Lambda$. 
For a limited range of the ``bifurcation parameter'' $\Lambda$ we find a
class of solutions which can be viewed as a continuation of the 
remarkable discrete family of particle--like solutions discovered by 
Bartnik and McKinnon (BK) for $\Lambda=0$ \cite{BMK}. 
In the vicinity of the origin, these solutions resemble the BK solitons.
However, the solutions are surrounded by a cosmological horizon and 
approach deSitter spacetime in the asymptotic region.
For each node number, $n$, these asymptotically deSitter solutions exist
only for sufficiently small cosmological constants,
$0 < \Lambda \leq \Llow(n)$, where we determine $\Llow(n)$ numerically.

When $\Lambda$ exceeds $\Llow(n)$ we obtain a different class of solutions,
for which the $2$--spheres (i.e., orbits belonging to the assumed 
SO$(3)$ symmetry) reach their maximal size {\em outside} the 
cosmological horizon. 
The position of the maximal sphere, $S^2_{\mbox{\scriptsize  max}}$, 
moves inwards as $\Lambda$ increases and approaches the horizon 
when $\Lambda$ tends to some special value $\Lmid(n)$, say. 
Outside $S^2_{\mbox{\scriptsize max}}$ a true singularity develops. 
This region resembles the interior of a black hole solution, 
whose singularity is also shielded by a horizon. 
For obvious reasons, we call these solutions {\em bag of gold}
configurations.

These bag of gold solutions continue to exist for 
$\Llow(n) < \Lambda < \Lmid(n)$, where the extremal sphere now lies 
{\em inside} the horizon. 
An interesting phenomenon occurs when $\Lambda$ reaches the upper 
limit $\Lhig(n)$, for which the singularity approaches the horizon. 
For $\Lambda=\Lhig(n)$ an {\em everywhere regular, spatially compact} 
solution exists for all $n$. 
In the special case where $n=1$ this is precisely the Einstein Universe 
with a constant energy density of the Yang--Mills field on $S^3$. 
This particular solution has repeatedly been rediscovered in the past 
\cite{Hosotani}. 
For higher node numbers, the spatial part of the manifold is a 
``squashed'' $3$--sphere, and the solutions can only be constructed 
numerically.

As is the case for the BK family, it would be valuable to 
have an existence proof for the compact solutions, probably along 
similar lines as presented in \cite{SY1}, \cite{SY3}. 
We would also like to mention Ref. \cite{maeda} on EYM solutions with
cosmological constant, 
which contains some partial results of the present paper.

A crucial issue is the question of stability
of the solutions presented in this paper. 
However, it turned out that this is a quite involved and subtle problem,
mainly for topological reasons. We shall therefore present this part of 
our investigation in an accompanying paper \cite{stable}.

This article is organized as follows: In the second and third sections we 
derive the basic equations and present some special solutions which can
be given in closed form. The fourth and fifth sections are
devoted to the asymptotically deSitter solutions and their
analytic extensions, respectively. The bag of gold configurations
are described in the sixth section. Finally, in the last section,
we discuss the globally regular, compact solutions.

\section{Basic Equations}

We consider an EYM model with cosmological constant $\Lambda$
and action
\be
S \, = \, - \frac{1}{4\pi}
\int \left[ \,
\frac{1}{4 G} \, \ast \, ({\cal R}-2\Lambda) \, + \, 
\frac{1}{2g^2} \, \mbox{tr} \, (F \wedge \ast F) \, \right] \, ,
\label{1:1}
\ee
where $G$ is Newton's constant and
$g$ denotes the gauge coupling constant. 
Since we restrict ourselves to
configurations with spherical symmetry, the
spacetime manifold $(M,\gtens)$ has (locally) the
structure of a warped product, $M = \tilde{M} \times_{R} S^{2}$, 
with metric
\be
\gtens \, = \, \gtiltens \, + \, R^{2} \, \ghattens \, .
\label{g-four}
\ee
Here, $\gtiltens$ and $\ghattens$ denote the metrics on $\tilde{M}$ 
and $S^{2}$, respectively, and $R$ is a function on $\tilde{M}$.
Throughout this paper, quantities referring to $(\tilde{M},\gtiltens)$ are 
endowed with a tilde and those for $(S^{2},\ghattens)$ with a hat. 
The Einstein tensor for warped product manifolds becomes \cite{BHS96}
\bea
G_{ab} & = & \frac{2}{R}  \left[
\gtil_{ab} \daltil R - \nabtil_{a} \nabtil_{b} R \right] 
\, + \, \frac{1}{R^{2}} \gtil_{ab} \left[
(dR | dR) - 1 \right] \, ,
\label{G-ab} \\
G_{Ab} & = & 0 \, ,
\label{G-Ab} \\
G_{AB} & = & 
R^{2} \ghat_{AB} \left[ \frac{1}{R} \daltil R - \frac{1}{2} 
\tilde{\cal R} \right] \, ,
\label{G-AB}
\eea
where $\tilde{\cal R}$ denotes the Ricci scalar of 
$(\tilde{M},\gtiltens)$.
(Small and capital Latin letters are used for indices on 
$(\tilde{M},\gtiltens)$ and $(S^{2},\ghattens)$, respectively;
$a,\ b,\ c\ = 0,\ 1$ and $A,\ B,\ C\ = 2,\ 3$.) 
With respect to the diagonal parametrization of the metric
$\gtiltens$, 
\be
\gtiltens \, = \, - \, e^{2a(t,\rho)} \, dt^{2} \, + \, 
e^{2b(t,\rho)} \, d \rho^{2} \, ,
\label{g-diag}
\ee
which we shall often use in this paper, the d'Alembertian
of a function $R$, say,
and the Ricci scalar on $(\tilde{M},\gtiltens)$ are
\be
\daltil R \, = \, e^{-(a+b)} \left[ \, (e^{a-b} R')' - 
(e^{b-a} \dot{R} ) \, \dot{} \, \right]                                     \label{dAl}
\ee
and
\be
\tilde{\cal R} \, = \, -2\, e^{-(a+b)} \left[ \, (e^{a-b} a')' - 
(e^{b-a} \dot{b} ) \, \dot{} \, \right] \, ,
\label{R-til}
\ee
respectively.

For $SU(2)$, the spherically symmetric 
gauge potential has the general form
\be
A \, = \, a \, \hat{\tau}_{\rho} \, + \, \varpi \, 
[\hat{\tau}_{\vartheta} d\vartheta + 
\hat{\tau}_{\varphi} \sin \vartheta d\varphi ] \, + \, 
(w-1) \,
[\hat{\tau}_{\varphi} d\vartheta - 
\hat{\tau}_{\vartheta} \sin \vartheta d\varphi ] \, ,
\label{1:3}
\ee
where $a = a_0 dt + a_1 d \rho$, and  
$a_{0}$, $a_{1}$, $w$ and $\varpi$ are functions on
$\tilde{M}$. Here 
$\hat{\tau}_{\rho}=n^{i}\tau^{i}/2$, 
$\hat{\tau}_{\vartheta}=\partial_{\vartheta}\hat{\tau}_{\rho}$,
$\hat{\tau}_{\varphi}=\partial_{\varphi}\hat{\tau}_{\rho}/\sin\vartheta$
and 
$n^{i}=(\sin\vartheta
\cos\varphi,\sin\vartheta\sin\varphi,\cos\vartheta)$, 
where $\tau^{i}$ are the Pauli matrices. 
In the static, purely magnetic case the choice
$a = \varpi= 0$ is compatible with the field equations. The 
gauge potential (\ref{1:3}) now reduces to 
\be
A \, = \, (w-1) \, [ \, \hat{\tau}_{\varphi} \, d\vartheta \, - \, 
\hat{\tau}_{\vartheta}\sin\vartheta \, d\varphi \,] \, .
\label{1:4}
\ee
In terms of $w$, the stress--energy 
tensor has the components
\bea
8 \pi g^2 T_{ab} & = & \frac{1}{R^{2}} \left[
w,_{a} w,_{b} \, - \, \gtil_{ab} \, 
\left( \frac{1}{2} (dw | dw) + \frac{V(w)}{4 \, R^{2}} \right) \right], 
\label{T-ab} \\
8 \pi g^2 T_{Ab} & = & 0 \, ,
\label{T-Ab} \\
8 \pi g^2 T_{AB} & = & \ghat_{AB} \, 
\frac{V(w)}{4 \, R^{2}} \, ,
\label{T-AB}
\eea
with $V(w) = (1-w^{2})^{2}$.

With respect to the parametrization (\ref{g-diag}) of the metric 
$\gtiltens$, the {\em static} field equations assume the form
\bea
- \, e^{-2 b} \, \left[
\mu'' + \mu' \, (\mu' - a' - b' ) \right] & = &
\kappa \; e^{-2 b} \, \frac{w'^2}{R^2} \, ,
\label{s-0}\\
\frac{1}{R^2} \, - \, 
e^{-2 b} \, \left[
\mu'' + \mu' \, ( 2 \mu' + a' - b' ) \right]
& = &
\kappa \; \frac{V(w)}{2 \, R^4} \, + \, \Lambda \, ,
\label{s-1}\\
\; \;
\frac{1}{R^2} \, + \, 
e^{-2 b} \, \left[
a'' + a' \, ( a' - b' ) - \mu'^2 \right]
& = &
\kappa \; \frac{V(w)}{R^4} \, ,
\label{s-2}
\eea
and
\be
e^{-(a+b)} \, (e^{a-b} \, w')' \, = \, \frac{1}{4 \, R^{2}} \, V,_{w} \, ,
\label{s-3}
\ee
where we have introduced $e^{\mu} \equiv R$ and where
eqs. (\ref{s-0}), (\ref{s-1}) and (\ref{s-2})
are the $\frac{1}{2} (00+11)$,
$\frac{1}{2} (00-11)$ and
$\frac{1}{2} (00+11-22-33)$ components of the
Einstein equations.
We also note that the (dimension--full) coupling constant 
$\kappa = 8 \pi G/ g^2$ can be absorbed by introducing
the dimensionless quantities 
$R/\sqrt{\kappa}$, $\rho/\sqrt{\kappa}$ and $\Lambda \kappa$.
(We shall often set $\kappa = 2$ in this paper, that is, we 
measure length, time and mass in units of
$[G g^2 c^{-4}]$, $[G^{1/2} g c^{-3}]$ and $[g^2 G^{-1}]$,
respectively; see \cite{JS81}.)

We shall use two gauges in this paper, depending on whether 
or not $R$ has a local maximum.
Considering solutions for which $R$ has no critical point, 
we can use Schwarzschild coordinates, that is, we 
are alowed to choose the gauge
\be
R(\rho) \, = \, \rho \, \equiv \, r \, .
\label{Schw-gauge}
\ee
It is then also convenient to introduce the functions
$N(r)$ and $\sigma(r)$, defined by
\be
N \, \equiv \, (dr | dr) \, = \, e^{-2b} \, ,
\; \; \; \; \; 
\sigma \, \equiv \, \sqrt{-\tilde{g}} \, = \, e^{a+b} \, .
\label{N-sig}
\ee
In terms of this parametrization, the static equations (\ref{s-0}), 
(\ref{s-1}) and (\ref{s-3}) become
\be
\sigma' \, = \, \kappa \, \frac{w'^2}{r} \, \sigma \, ,
\label{ssw-0}
\ee
\be
m' \, = \, \frac{\kappa}{2} \, \left[
N \, w'^2 \, + \, \frac{V(w)}{2 \, r^2} \right] \, ,
\label{ssw-1}
\ee
\be
N \, w'' \, + \, \frac{w'}{r} \, \left[
\frac{2 m}{r} - \frac{2}{3} \Lambda r^2 - \kappa \frac{V}{2 r^2}
\right] \, = \, \frac{V,_w}{4 \, r^2} \, ,
\label{ssw-3}
\ee
where a dash denotes the derivative with respect to $r$.
Here we have already used eq. (\ref{ssw-0}) 
in the second and the third equations, in order to eliminate 
the metric function $\sigma$. 
The function $m(r)$ is defined by the relation
\be
N(r) \, \equiv \, 1 \, - \, \frac{2m(r)}{r} \, - \, \frac{\Lambda}{3} 
\, r^2 \, .
\label{1:6}
\ee

When considering solutions for which 
$R$ develops a local extremum, we  
use the gauge $a+b = 0$, that is, we parametrize the static
metric by the two functions $R(\rho)$ and $Q(\rho)$, where
\be
Q(\rho) \, \equiv \, e^{2a} \, = \, e^{-2b} \, . 
\label{def-Q}
\ee
The static field equations 
(\ref{s-0}), (\ref{s-2}) and (\ref{s-3})
then assume the form
\be
R'' \, = \, - \, \kappa \; \frac{w'^2}{R} \, , 
\label{ssq-0}
\ee
\be
Q'' \, = \, 2 Q (\frac{R'}{R})^2 \, - \frac{2}{R^2}
\, + \, \kappa \, \frac{2 V(w)}{R^4} \, ,
\label{ssq-2}
\ee
\be
\left( Q \, w' \right)' \, = \, \frac{V,_w}{4 \, R^2} \, ,
\label{ssq-3}
\ee
where now $Q' \equiv dQ/d\rho$, etc.
Using eq. (\ref{ssq-0}), the remaining equation (\ref{s-1})
becomes a first integral,
\be
(Q \, R)' \, R' \, = \, \kappa \, \left(
Q w'^2 \, - \, \frac{V(w)}{2 \, R^2} \right)
 \, + \, 1 \, - \, \Lambda R^2 \, .
\label{ssq-1}
\ee
It is clear that this coordinate system is also 
suited to discuss solutions for which
$R$ has no critical points.
However, for obvious reasons, we prefer to use the
familiar parametrization (\ref{Schw-gauge}), (\ref{N-sig}) 
in those cases.

\section{Special Solutions}

Before we present a classification of the static configurations,
we consider some special solutions which can
be given in closed form.

First, for $R(\rho) = \rho$ and constant Yang--Mills amplitude,
we find from eqs. (\ref{ssq-0})-(\ref{ssq-1}) above: 
\be
R(\rho) \, = \, \rho,\ \ \ \
w(\rho) \, = \, 0,\pm 1,\ \ \ \
Q(\rho) \, = \, 1 \, - \, \frac{2M}{\rho} \, + \, \kappa \,
\frac{V(w)}{2 \rho^{2}} \, - \, \frac{\Lambda}{3}\rho^{2},          \label{1:11}
\ee
with $M$ being a constant of integration.
For $w=0$ ($V=1$) this solution corresponds to the
Reissner--Nordstr\"om--deSitter Universe with unit magnetic charge,
whereas we obtain the Schwarzschild--deSitter solution 
for $w=\pm 1$ ($V=0$).  

Next, we consider solutions for which both $R(\rho)$ and
$w(\rho)$ are constants. For $V(w) = 0$ one easily finds
\be                
R(\rho)\, = \, \frac{1}{\sqrt{\Lambda}},\ \ \ \ 
w(\rho) \, = \, \pm 1,\ \ \ \
Q(\rho) \, = \, - \, \Lambda \rho^{2} \, + \, A \rho \, + \, B \, ,
\label{1:12}
\ee
which corresponds to the $H^{2}\times S^{2}$
Nariai solution \cite{nariai}. 
(Here $A$ and $B$ are constants of integration.)
If $w=0$, we find for sufficiently small 
values of the cosmological constant, $\Lambda \leq (2 \kappa)^{-1}$,
\be
R^2(\rho) \, = \, \frac{1\pm\sqrt{1-2 \kappa \Lambda}}{2\Lambda} \, ,\ \ \ 
w(\rho) \, = \, 0,\ \ \ \
Q(\rho) \, = \, \frac{1}{R^2} (\frac{\kappa}{R^2}-1) \rho^2 \, + \, 
A \rho \, + \, B \, .
\label{1:13}
\ee
In the limit of vanishing $\Lambda$ the solution
with the lower sign reduces to the magnetic 
Robinson--Bertotti Universe (with $R^2 = \frac{1}{2} \kappa$). 

Finally, there exists a solution for which the components of the
stress--energy tensor assume constant values without $w(\rho)$ 
being a constant. This is possible only for the special 
value $\Lambda=3/(2 \kappa)$. In fact,
\be
R(\rho) \, = \, \sqrt{\kappa} \, \sin({\rho\over \sqrt{\kappa}}),\ \ \
w(\rho) \, = \, \cos({\rho\over \sqrt{\kappa}}),\ \ \
Q(\rho)\, = \, 1 
\label{1:14}
\ee
describes the static Einstein Universe. 

The above examples indicate that the 
qualitative behavior of the static solutions
to eqs. (\ref{s-0})-(\ref{s-3})
crucially depends on the value of the cosmological 
constant. In the following, we shall
present a classification of these solutions 
in terms of $\Lambda$ and the node number of $w$. 

\section{Asymptotically deSitter Solutions}

We start our numerical investigation by considering 
small values of $\Lambda$.
For $\Lambda=0$ the regular, 
asymptotically flat solutions of the
EYM equations were found by Bartnik and McKinnon
in 1988 \cite{BMK} and, since then, have been subject to numerous studies
(see, e.g. \cite{BHS96}, \cite{mavr}, \cite{SY3} and references therein).
Each solution has a typical size, $R_{n}$, where $n$ is the
number of nodes of the YM amplitude $w$.
In the region $R>R_{n}$ the energy density 
of the Yang--Mills field decays rapidly, and the metric 
approaches the vacuum Schwarzschild metric.

For small values of the cosmological constant,
$\Lambda\ll 1/R^{2}_{n}$, the contribution 
$\Lambda R^{2}$ to the energy density
is negligible. 
For $R<R_{n}$, one therefore expects that 
the solutions do not considerably deviate
from the BK solutions. 
In the region $r>R_{n}$, however, the effect of 
$\Lambda$ becomes significant,
which suggests that the metric
approaches the deSitter metric.
Hence -- for sufficiently small values of the cosmological 
constant -- the solutions are expected to resemble 
the regular BK solitons, which are surrounded by a cosmological 
horizon at $R\sim 1/\sqrt{\Lambda}$ and approach
the deSitter geometry in the asymptotic region.

\begin{figure}
\epsfxsize=10cm
\centerline{\epsffile{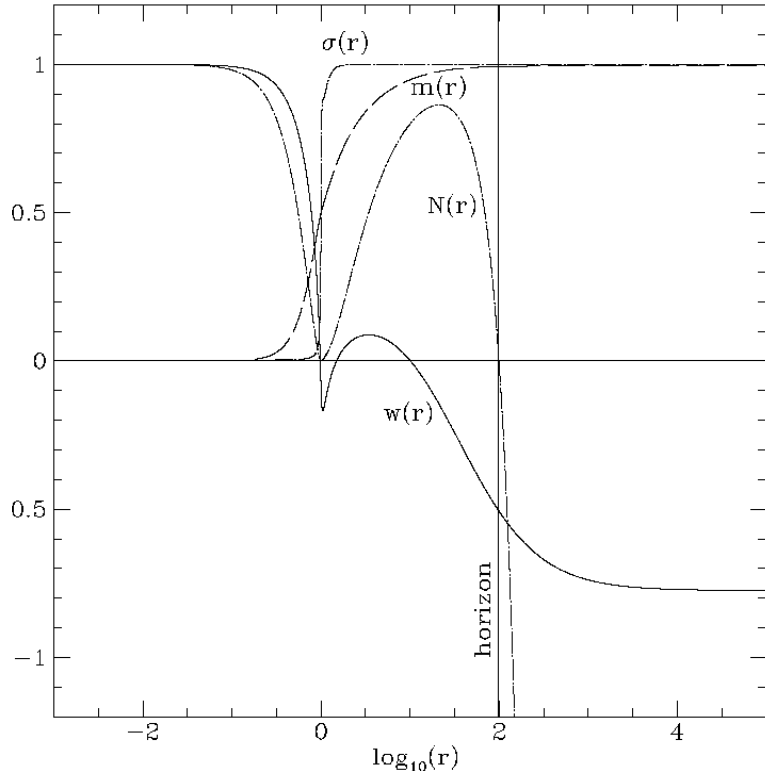}}
\caption{Asymptotically deSitter solution with
$\Lambda=3\times 10^{-4}$ and $n=3$. 
For this solution one finds 
$b=0.6998$, $r_{h}=98.99$, 
$w_{h}=-0.505$, $w_{\infty}=-0.774$, 
$M=m(\infty)=0.994$, $a=-37$,
$\sigma(0)=2\times 10^{-3}$, $\sigma(r_{h})=0.99999$ and
$N_{min}=1.7\times 10^{-3}$.}
\label{Fig.1}
\end{figure}

The numerical analysis of eqs. (\ref{ssw-0})-(\ref{ssw-3}) 
confirms these expectations.
In order to find numerical solutions, we need the formal power
series expansions of the equations (\ref{ssw-1}) and (\ref{ssw-3})
in the vicinity of the origin, $r=0$, 
the cosmological horizon, 
$r=r_{h}$, and for $r \rightarrow \infty$.
In the vicinity of the origin, the regular 
solutions behave as follows ($\kappa = 2$):  
\be 
w \, = \, 1-br^2+{\cal O}(r^4),\ \ \ \ \ \
N \, = \, 1-(4b^2+{\Lambda\over 3})r^2+{\cal O}(r^4) \, .
\label{2:1}
\ee
Near the horizon, defined by $N(r_h) = 0$, we find
with $x=r-r_{h}$:
\be
w \, = w_h + w'_{h} \, x + {\cal O}(x^2),\ \ \ \ \ \
N \, = \, N'_{h} \, x + {\cal O}(x^2) \, ,
\label{2:2}
\ee
where
\be
N'_{h} \, = \, \frac{1}{r_{h}}(1 - \frac{V(w_h)}{r_h^2} - 
\Lambda r_{h}^{2})\ < 0
\ \ \ {\rm and} \ \ \
w'_{h} \, = \, \frac{V,_w(w_h)}{4 r^2_h N'_h} \, .
\label{hor-exp}
\ee
Finally, in the asymptotic regime, $r \rightarrow \infty$, we have
\bdm
w \, = \, w_{\infty} \, + \, \frac{a}{r} \, - \, 
\frac{3 V,_w(w_{\infty})}{8 \, \Lambda}  \frac{1}{r^2} \, + \, 
{\cal O}(\frac{1}{r^3}) \, ,
\edm
\be
N\, = \, 1 \, - \, \frac{2M}{r} \, - \, \frac{\Lambda}{3}\, r^{2}
\, + \, [V(w_{\infty}) - \frac{2}{3} \Lambda a^2 ] \frac{1}{r^2}
\, + \,  {\cal O}(\frac{1}{r^{3}}) \, .
\label{2:3}
\ee
Here, $b$, $r_{h}$, $W_{h}$, $W_{\infty}$, $M$ and $a$ are
six ``shooting'' parameters.

In order to obtain numerical solutions to the static equations one starts the integration with the expansions 
(\ref{2:1}) and (\ref{2:2}) and tries to match the
functions $w$, $w'$ and $N$ at some intermediate 
point between the origin and the horizon. 
The three matching conditions then fix the 
values of the parameters $b$, $r_{h}$ and $w_{h}$ 
appearing in eqs. (\ref{2:1}) and 
(\ref{2:2}). Subsequently, one uses the remaining
three parameters in eq. (\ref{2:3}) to match the solutions obtained 
from numerically integrating between 
the horizon and infinity. Finally, the remaining
metric function $\sigma$ is obtained from eq. (\ref{ssw-0}), where
$\sigma$ behaves like
\be
\sigma \, = \, \sigma(0)+{\cal O}(r^{2}),\ \ \  
\sigma \, = \, \sigma(r_{h})+{\cal O}(r-r_{h}),\ \ \
\sigma \, = \, 1+{\cal O}(\frac{1}{r^{4}}),  \ \ \
\label{2:4}
\ee
in the vicinity of the origin, the horizon and infinity, respectively (see Fig.1).

The numerical procedure yields the following result:
For each fixed value of $\Lambda\ll 1$ we recover a family 
of solutions which correspond to the $\Lambda=0$
BK solitons. 
Each solution is characterized by the value of $\Lambda$ and the 
number, $n$,  of nodes of $w$ inside the cosmological horizon.
Outside the horizon $w$ tends to a constant value, $w_{\infty}$, say. 
Since $w_{\infty}  \neq \pm 1$, the YM field gives rise to the
magnetic charge 
${\cal P} = [2 \, \mbox{tr}(P \cdot P)]^{1/2} = w^{2}_{\infty}-1$,  
where
\be
P \,  = \, \frac{1}{4\pi} \oint_{S^{2}} \, F \, ,
\label{2:5}
\ee
and where the integration is performed over the $2$--sphere
at spatial infinity. Here we have used eq. (\ref{1:4}) and
$F = dA + A \wedge A$ to obtain 
$F = (w^2-1) \hat{\tau}_{\rho} d \Omega + (w-1)^{-1}dw \wedge A$.
(It is worthwhile recalling that the solutions with $\Lambda=0$  
have vanishing magnetic charge.) 
The metric asymptotically approaches the Reissner--Nordstr\"om--deSitter 
metric with effective charge $[P^2 - 2 \Lambda a^2 /3]^{1/2}$;
see eq.(\ref{2:3}).

\section{Analytic Extensions}

\begin{figure}
\epsfxsize=10cm
\centerline{\epsffile{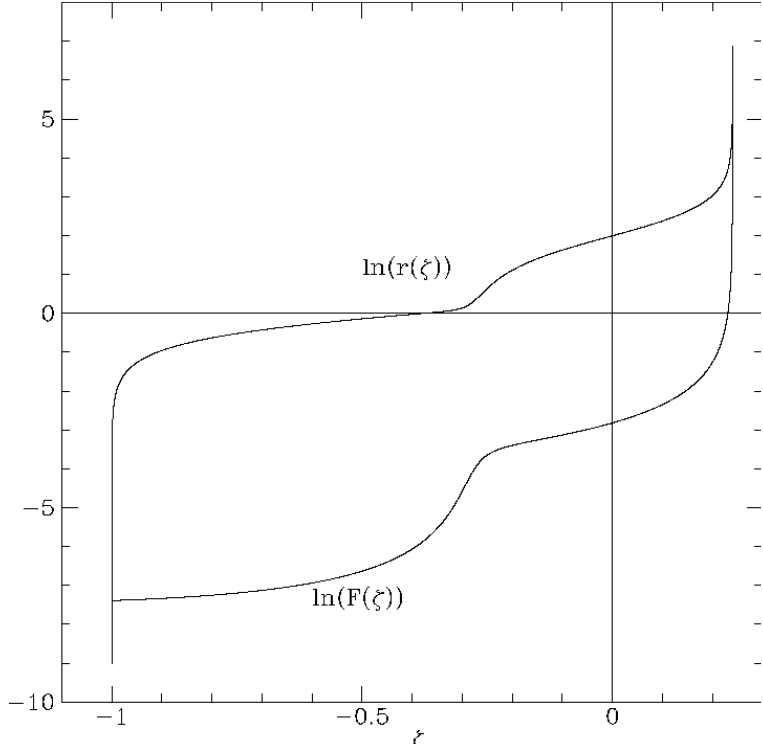}}
\caption{The functions $r(\zeta)$ and $F(\zeta)$ 
for the asymptotically deSitter solution with
$\Lambda=3\times 10^{-4}$, $n=2$. For this solution one has
$F(0)=6\times 10^{-4}$ and
$\zeta_{\infty}=0.24$.}
\label{Fig.2}
\end{figure}

In this section we construct the analytic extension 
for a generic metric of the above type.
Our first goal is to write the metric $\gtiltens$ 
in conformally flat form, such that the spacetime metric
becomes
\be
\gtens \, = \, \sigma^{2} \,  N \, (dt^{2}-d\chi^{2}) \, - 
\, r^{2}d\Omega^{2} \, .
\label{3:4}
\ee
In order to do so, we need the following essential properties
of the solutions discussed above: 
Both $N$ and $\sigma$
are smooth functions, where $\sigma(r)$ is bounded and
everywhere positive. The metric function $N(r)$ 
is subject to the boundary conditions
$N(0)=1$ and $N \rightarrow -c^{2} r^{2}$ 
as $r\rightarrow\infty$. Moreover, $N(r)$ changes sign 
exactly once, namely at the horizon, $N(r_{h})=0$.
By virtue of these properties, the 
new radial coordinate $\chi$,
\be
\chi(r)= \int_{0}^{r}\frac{d\bar{r}}{\sigma N} \, , \ \ r<r_{h},
\ \ \ {\rm and}\ \ \
\chi(r)= \chi_{\infty} - \int_{r}^{\infty}\frac{d\bar{r}}{\sigma N},
\ \ r>r_{h}
\label{3:2}
\ee
increases from zero to infinity as $r$ runs from
zero to $r_{h}$, and then decreases from infinity to 
$\chi_{\infty}$ as $r$ grows from $r_{h}$ to infinity. 
The constant $\chi_{\infty}$ is fixed by
considering the expansion of the above integrals
in the vicinity of the horizon,
\be
\chi=-\frac{1}{2\eta}\ln|r-r_{h}|+C+{\cal O}(r-r_{h}) \, ,
\ \ \ |r-r_{h}|\ll 1 \, , 
\label{3:3}
\ee
and requiring that the constant $C$ has the same
value in both cases.
Here we have also introduced
the quantity $\eta$,
\be
\eta\, = \, - \, \frac{1}{2} \, 
\sigma \, N'\left|_{r=r_{h}}\, > \, 0 \, , \right.
\label{3:1}
\ee
which does not vanish for a regular horizon.
With respect to $\chi$, the metric now assumes the
desired form (\ref{3:4}) which, in a neighborhood
of the horizon, becomes
\be
\gtens \,  = \, \pm\sigma(r_{h}) \, \eta \, 
e^{-2\eta\chi} \,  [1+{\cal O}(e^{-2\eta\chi})] \,
(dt^{2}-d\chi^{2}) \, - \, r^{2} d\Omega^{2} \, ,
\label{3:5}
\ee
where the plus and minus signs refer to the regions
$r<r_{h}$ and $r>r_{h}$, respectively. 

\begin{figure}
\epsfxsize=10cm
\centerline{\epsffile{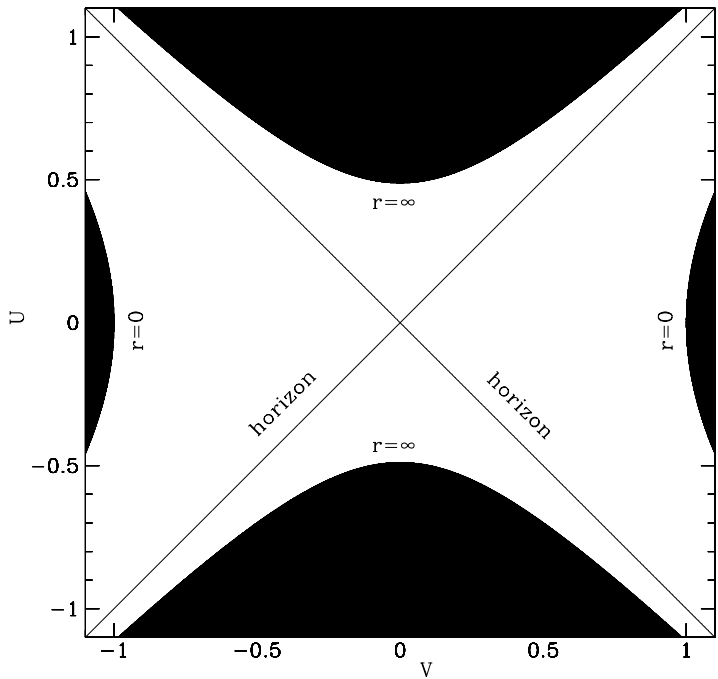}}
\caption{Spacetime diagram 
for the asymptotically deSitter solution with
$\Lambda=3\times 10^{-4}$, $n=2$.  
}
\label{Fig.3}
\end{figure}

Next, we note that $\zeta(r)$, defined by
\be
\zeta(r) \, = \, - 
e^{-2\eta\chi(r)},\ \ \ r<r_{h} \, ,\ \ \ \ \ \ \   \ {\rm and}\ \ \ \
\zeta(r) \, = \, e^{-2\eta\chi(r)},\ \ \ r>r_{h} \, ,
\label{3:6}
\ee
is a monotonically increasing function of $r$ with
$\zeta (0)=-1$, $\zeta(r_{h})=0$ and 
$\zeta \rightarrow \exp (-2\eta\chi_{\infty})>0$
as $r \rightarrow \infty$.
Hence, the inverse function, $r(\zeta)$, is
well--defined and the function $F(\zeta)$,  
\be
F(\zeta) \, = \, - \, \frac{\eta^{2}}{\zeta} \, \sigma^{2}
(r(\zeta)) \, N(r(\zeta)) \, ,
\label{3:7}
\ee
is therefore smooth and everywhere positive.
As usual, one finally passes from
the coordinates $(t,\chi)$ to the new coordinates $(U,V)$, where
\bea
U & = & e^{-\eta\chi}\sinh\eta\chi,\ \ \ \ \
V=e^{-\eta\chi}\cosh\eta\chi,\ \ \ \ r<r_{h} \, ,
\nonumber\\
U & = & e^{-\eta\chi}\cosh\eta\chi,\ \ \ \ \
V=e^{-\eta\chi}\sinh\eta\chi,\ \ \ \ r>r_{h} \, .
\label{3:8}
\eea
The analytically extended metric eventually becomes
\be
\gtens \, = \, 
F(\zeta) \, (dU^{2}-dV^{2}) \, - \, r^{2}(\zeta) \, d \Omega^{2}, 
\label{3:9}
\ee
where $\zeta=\zeta(U,V)=U^{2}-V^{2}$. The two functions
$F(\zeta)$ and $r(\zeta)$ can be determined numerically (see Fig.2).
For the deSitter solution one easily finds
\be
F(\zeta) \, = \, \frac{4}{(1-\zeta)^{2}},\ \ \ \
r(\zeta) \, = \, \frac{1+\zeta}{1-\zeta},\ \ \ \ \zeta\in [-1,1]. \label{3:10}
\ee 

The spacetime diagram in coordinates $(U,V)$ is displayed in 
Fig.3. The spacetime manifold corresponds 
to the region $U^{2}-V^{2}\in[-1,\zeta_{\infty}]$. The 
qualitative features of the diagram are identical with those of
the deSitter solution.  

\begin{figure}
\epsfxsize=10cm
\centerline{\epsffile{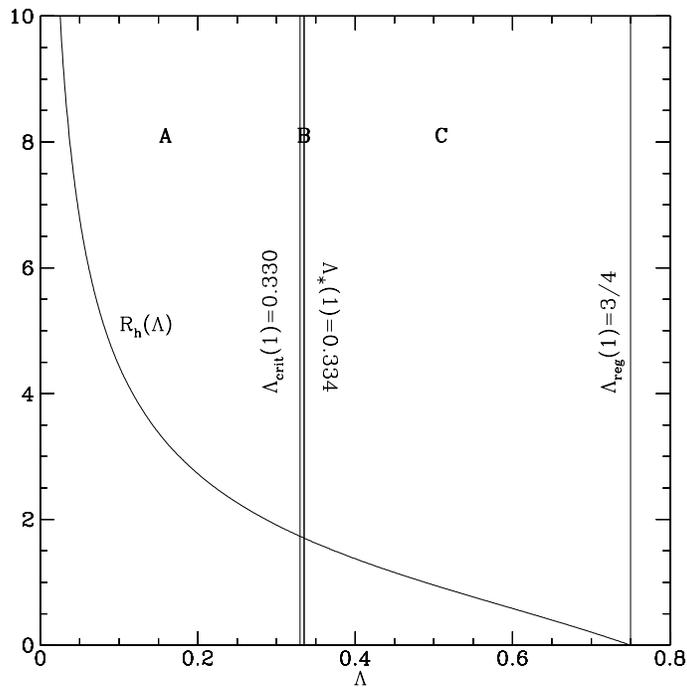}}
\caption{The horizon radius $R_{h}$ versus the cosmological
constant $\Lambda$ for the $n=1$ EYM solutions. }
\label{Fig.4}
\end{figure}

\section{Bag of Gold Solutions}

The asymptotically deSitter solutions described above exist only for 
sufficiently small values of the cosmological constant: 
For each fixed value of the node 
parameter $n$, there exists a maximal value $\Llow(n)$, say, 
beyond which the numerical analysis breaks down.

Solutions which belong to larger values of $\Lambda$ 
exhibit a local extremum of $R$ and cannot be
obtained in Schwarzschild coordinates. 
We therefore pass to a parametrization
of the metric for which $R(\rho)$ is a dynamical function and
choose the gauge $e^{2a} = e^{-2b} \equiv Q(\rho)$; 
see eq. (\ref{def-Q}). 

Equations (\ref{ssq-0})-(\ref{ssq-1}) yield  
the formal power series at the origin ($\kappa = 2$), 
\be
w \, = \, 1 - b \, \rho^{2} + {\cal O}(\rho^{4}),\ \ \ \
R \, = \, \rho + {\cal O}(\rho^{5}),\ \ \ \ 
Q \, = \, 1+(4b^{2}-\frac{\Lambda}{3})\rho^{2}+{\cal O}(\rho^{4}),
\label{power-orig-gold}
\ee
where $b$ is the only free parameter.
The numerical integration shows that
$Q(\rho)$ develops a zero at some  
$\rho=\rho_{h}(b,\Lambda)$, 
indicating the presence of a horizon.  
Requiring that all curvature
invariants remain finite at the horizon 
yields
\be
\lim_{\rho\rightarrow\rho_{h}} \sqrt{Q} \, w' \, = \, 0 \, . 
\label{4:2}
\ee

\begin{figure}
\epsfxsize=10cm
\centerline{\epsffile{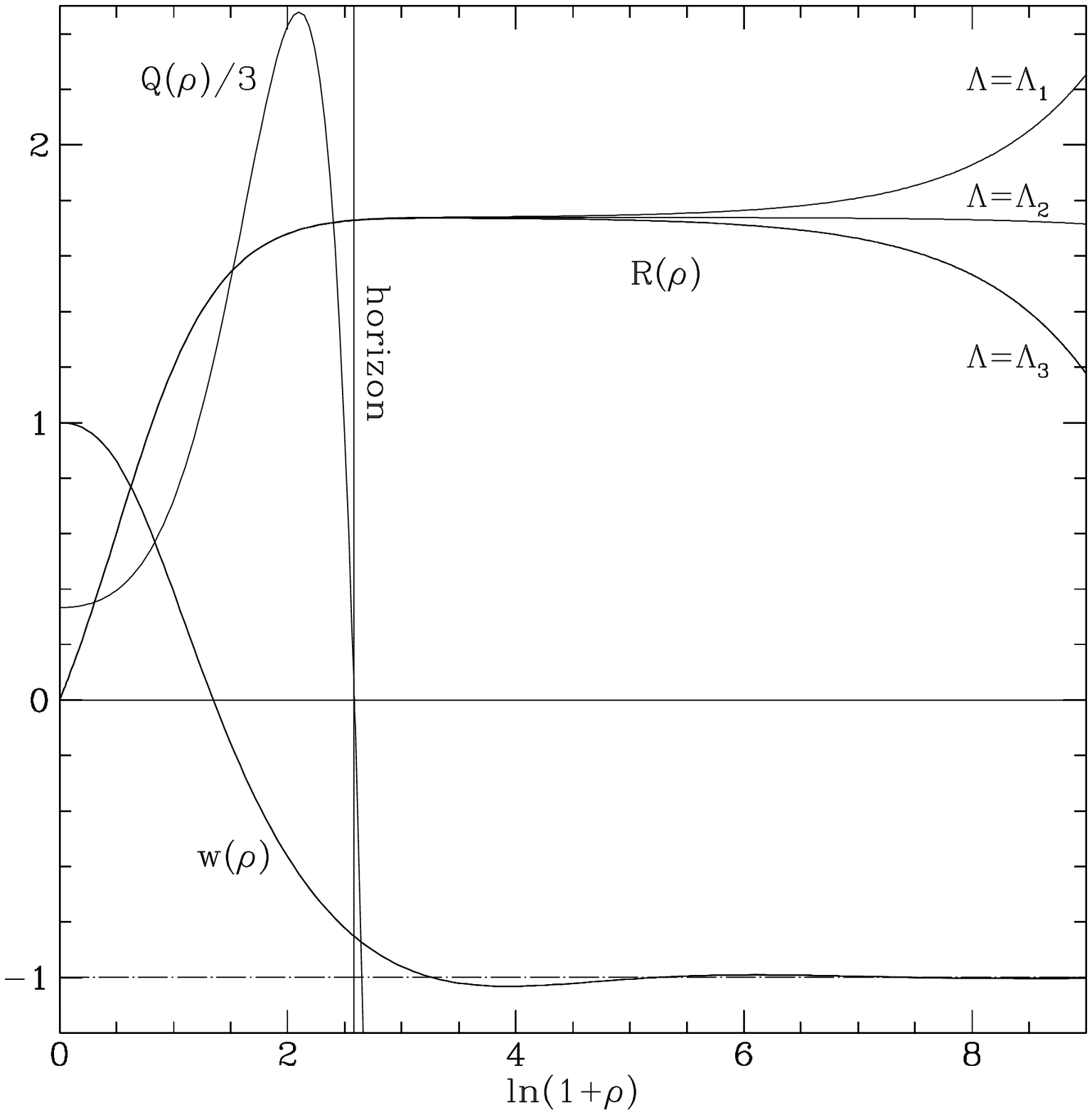}}
\caption{Change of the topology of the EYM solutions. 
The solution with $\Lambda=\Lambda_{1}=0.3304$ is asymptotically deSitter, 
whereas the one with $\Lambda=\Lambda_{3}=0.3306$ is of the 
bag of gold type. 
The value $\Lambda=\Lambda_{2}=0.3305$ is very close to 
$\Lambda_{{\rm crit}}$. 
The functions $Q(\rho)$ and $w(\rho)$ for the three solutions are
almost identical. }
\label{Fig.5}
\end{figure}

As a consequence of this condition we obtain
a family of solutions between the origin and
the horizon, which are parametrized by 
a discrete set of values $b_{n}(\Lambda)$, where 
$n$ is the number of nodes. The 
parameters $w_{h}$, $R_{h}$ and $R'_{h}$ 
entering the power series at the horizon,
\be
w=w_{h}+w'_{h} x + {\cal O}(x^{2}),\ \ \ \
Q=Q'_{h} x+ {\cal O}(x^{2}),\ \ \ \
R=R_{h}+R'_{h} x+O(x^{2}),\ \ \ \
\label{power-hor-gold}
\ee
are therefore fixed, once $b_n(\Lambda)$ is known.
Here, $x = \rho - \rho_{h}$, and 
$Q'_{h}$ and $w'_{h}$ are given in terms of
$w_{h}$, $R_{h}$ and $R'_{h}$:
\be
Q'_{h} \, = \, \frac{1}{R_{h} \, R'_{h}} 
\left( 1 - \frac{V(w_{h})}{R_{h}^{2}} - \Lambda R_{h}^{2}\right) \, ,
\ \ \ \
w'_{h} \, = \, \frac{V,_{w}(w_{h})}{4 R_{h}^{2} \, Q'_{h}} \, .
\label{4:3}
\ee
Finally, we use this expansions to extend the solution beyond 
the horizon. The advantage of this procedure is that it essentially uses only 
{\em one} shooting parameter, $b$; the remaining parameters
are then iteratively determined.

The numerical analysis reveals the following picture:
For each value of the node number,
the horizon radius $R_{h}$
decreases monotonically with increasing
values of $\Lambda$, where 
$R_{h}\rightarrow\infty$ for $\Lambda\rightarrow 0$
and $R_{h}\rightarrow 0$ for $\Lambda\rightarrow \Lhig(n)$ (see Fig.4). 
The limiting value $\Lhig(n)$, for which the horizon shrinks to zero, 
decreases with growing node number $n$, where
$\Lhig(1)=3/4$ and  $\Lhig(\infty)=1/4$. 
\begin{figure}
\epsfxsize=10cm
\centerline{\epsffile{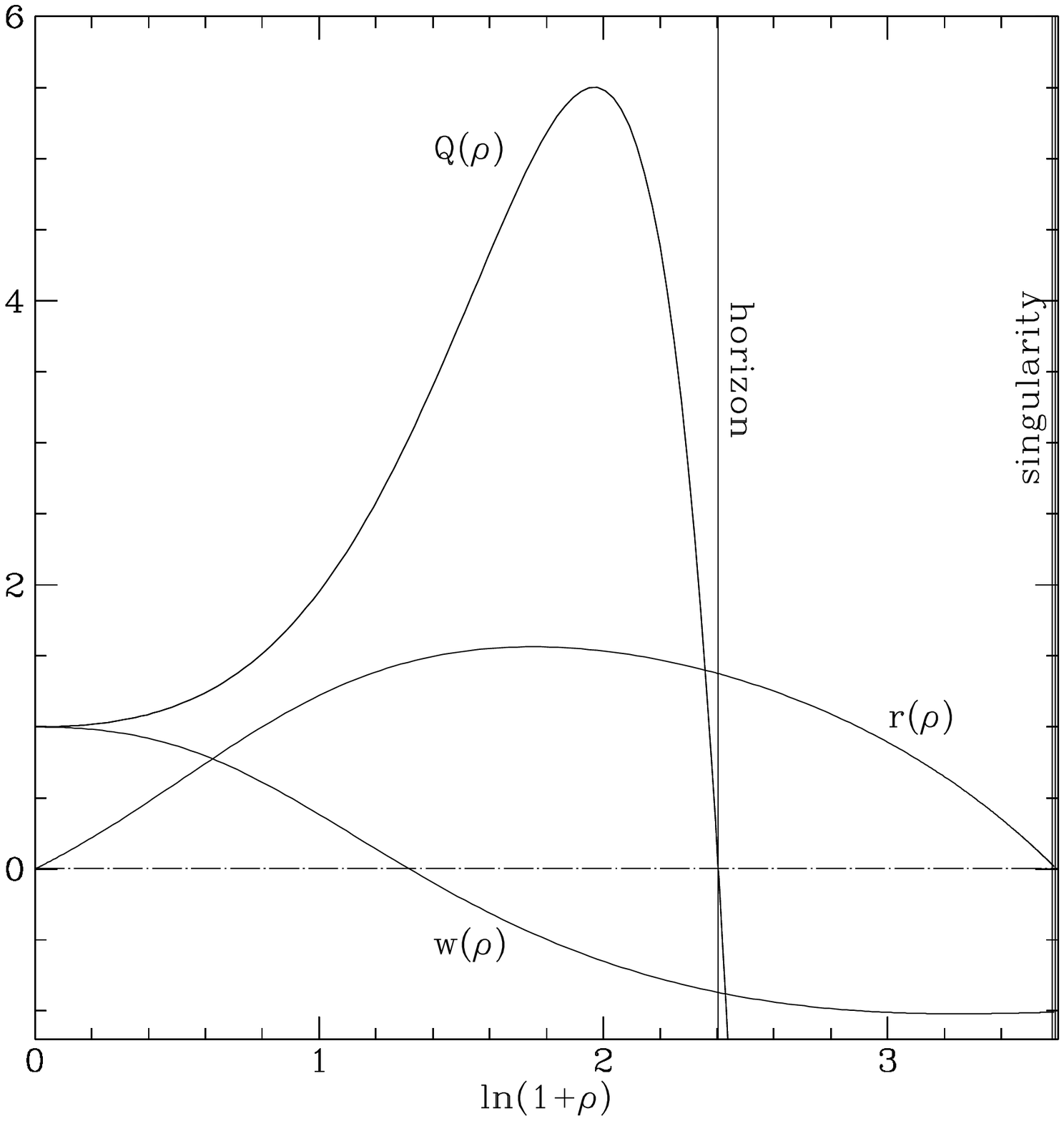}}
\caption{The bag of gold solution with $\Lambda=0.4$ and $n=1$. }
\label{Fig.6}
\end{figure}

Depending on the position of the maximum of $R$,
one finds three qualitatively different classes of
solutions, corresponding to the following subdivision
of the interval $(0,\Lhig(n))$ (see Fig.4). \\
{\bf{(A)}} $0 < \Lambda < \Llow(n)$:
These are the deSitter like solutions discussed earlier.
The function $R(\rho)$ has no critical points for finite
values of $\rho$. In the asymptotic regime, $R'(\rho)$ behaves like
\be
R'(\rho)=R'_{\infty}+{\cal O}(\frac{1}{\rho}) \ \ \ \ \ \ {\rm as}\ \ \ 
\rho\rightarrow\infty \, ,
\label{4:4}
\ee
where the constant $R'_{\infty}$ decreases with growing values of
$\Lambda$ and vanishes for $\Llow(n)$. Thus, for 
$\Lambda\rightarrow\Lambda_{{\rm crit}}(n)$,
$R(\rho)$ develops an ``extremum at infinity''.
The numerical analysis (see Fig.5) suggests that for   
$\Lambda =\Lambda_{{\rm crit}}(n)$, $R(\rho)$ asymptotically approaches
a constant value, $R(\infty)=1/\sqrt{\Lambda_{{\rm crit}}(n)}$, whereas 
$w(\rho)$ tends to $\pm 1$; such that for  
$\rho\rightarrow\infty$ the solutions coincide  
with the Nariai solution (\ref{1:12}). The topology of the solutions
therefore changes for $\Lambda =\Lambda_{{\rm crit}}(n)$
(for the solution with one node one has $\Llow(1)=0.3305$).\\
{\bf{(B)}} $\Llow(n) < \Lambda < \Lmid(n)$:
For these values of $\Lambda$ 
the function $R(\rho)$ develops a maximum for a finite
value $\rho_{e}$ {\it outside\/} the horizon,
$\rho_h < \rho_{e} < \infty$.
Since $R'' \leq 0$ (see eq. (\ref{ssq-0})), 
$R(\rho)$ decreases for 
$\rho > \rho_{e}$ and becomes zero at some finite
value $\rho_{sing}$, say.
In fact, the metric function $Q$ diverges as 
$\rho \rightarrow \rho_{sing}$,
indicating that the geometry becomes singular.
(For the solution with one node one finds
$\Lmid(1) = 0.334$.)\\ 
{\bf{(C)}} $\Lmid(n) < \Lambda < \Lhig(n)$:
The behavior is similar to case (B).
However, now $R$ reaches the maximal value
{\it inside} the horizon, 
$\rho_{e} < \rho_h$. Since $\Lambda_{{\rm reg}}(n)$
is the maximal value for which the solutions exhibit a horizon,
$R$ vanishes still outside the horizon, that is,
$\rho_{sing} \geq \rho_{h}$. Again, $Q$ is unbounded for
$\rho = \rho_{sing}$ (see Fig.6).

We call the solutions which exhibit a horizon and for which
$R$ develops a second zero outside the
horizon bag of gold solutions.

\section{Compact Regular Solutions}

Until now we have restricted our attention to
solutions which develop a horizon.
A new and interesting type of solutions 
is obtained in the limit $\Lambda\rightarrow\Lhig(n)$,
where the horizon and the singularity merge,
$\rho_h\rightarrow\rho_{sing}$. In this limit,
that is for $\Lambda = \Lhig(n)$, 
the geometry turns out to 
be everywhere regular, in particular
at both zeros of $R$. Moreover,
the points where $R$ assumes its maximal value, $\rho_e$,
lies precisely between these zeros and the spatial geometry
is symmetric with respect to $\rho_e$.
Since, in this case,  
the manifold has the topology of $I \! \! R \times S^{3}$,
the zeros of $R$ and the $2$--sphere 
$\rho = \rho_e$ will be called the 
north pole, the south pole and the equator, respectively.

For each node number $n$, there exists precisely one
value of the cosmological constant, $\Lambda = \Lhig(n)$,
for which one obtains {\it compact} solutions of the above kind.
For $n=1$, the solution is the static Einstein Universe
with $\Lambda_{\rm reg}(1) = 3/(2 \kappa)$, already 
presented in the second section:
\be
R(\rho) \, = \, \sqrt{\kappa} \, \sin({\rho\over \sqrt{\kappa}}),\ \ \
w(\rho) \, = \, \cos({\rho\over \sqrt{\kappa}}),\ \ \
Q(\rho)\, = \, 1 \, .
\label{1:14-bis}
\ee

The regular solutions with higher node numbers
are obtained in the limit $\Lambda\rightarrow\Lhig(n)$
from the corresponding bag of gold solutions.
An alternative method, which takes advantage of the
reflection symmetry with respect to the 
equator, is to integrate the field equations
on the ``northern hemisphere'', say.
In order to do so, one has to impose
boundary conditions at the pole (i.e., the origin,
$\rho = 0$) and the equator ($\rho = \rho_{e})$.
The solutions are then obtained by matching the
numerical integrations from the pole
and the equator.

\begin{figure}
\epsfxsize=10cm
\centerline{\epsffile{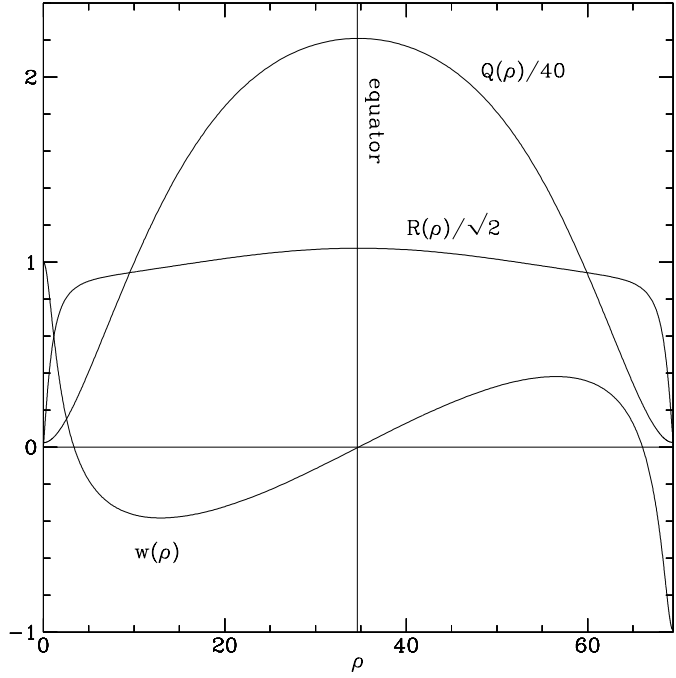}}
\caption{The $n=3$ compact solution.}
\label{Fig.7}
\end{figure}

The formal power series at the origin
involve one ``shooting'' parameter, $b$, and 
were given in eq. (\ref{power-orig-gold}).
In order to obtain the series expansions
in the vicinity of the equator, we
have to distinguish two cases: Depending
on whether the gauge field amplitude
$w(\rho)$ is antisymmetric or symmetric
with respect to $\rho_{e}$,
the regular compact solutions will be called 
odd ($w_{e} = 0$) or even ($w'_{e} = 0$), respectively.\\
{\bf{(i)}} $w_{e} = 0$:
For the odd configurations one finds with $x = \rho - \rho_{e}$
($\kappa =2$)
\bdm
R \, = \, R_{e} + \frac{1}{2}R''_{e} \, x^{2} + {\cal O}(x^{4}) \, , 
\; \; \; \; \; 
Q \, = \, Q_{e} + \frac{1}{2}Q''_{e} \, x^{2} + {\cal O}(x^{4}) \, , 
\edm
\be
w \, = \, w'_{e} \, x  \, + \, {\cal O}(x^{3}) , 
\label{eq-exp}
\ee
where the field equations (\ref{ssq-0})-(\ref{ssq-1}) imply that
$w'_{e}$, $R''_{e}$ and $Q''_{e}$ are given in terms of 
$R_{e }$ and $Q_{e}$,
\bdm
w'^{2}_{e} \, = \, \frac{1}{2Q_{e}} (R_{e}^{-2}+\Lambda R^{2} -1) , 
\, \, \, \, \, \, 
R''_{e} \, = \, -\frac{2}{R_{e}} w'^{2}_{e} ,
\, \, \, \, \, \, 
Q''_{e} \, = \, \frac{2}{R_{e}^{4}}(2 - R_{e}^{2}) \, . 
\edm
{\bf{(ii)}} $w'_{e} = 0$: 
For solutions with even Yang--Mills amplitude we have
\bdm
R \, = \, R_{e} + \frac{1}{4!} R^{(4)}_{e} \, x^{4} + {\cal O}(x^{6}) \, , 
\; \; \; \; \; 
Q \, = \, Q_{e} + \frac{1}{2}Q''_{e} \, x^{2} + {\cal O}(x^{4}) \, , 
\edm
\be
w \, = \, w_{e}  \, + \,  \frac{1}{2} w''_{e} \, x^{2} \, + \, 
{\cal O}(x^{4}) \, .
\label{eq-exp:1}
\ee
As before, the only free parameters are $R_{e }$ and $Q_{e}$.
In terms of these, $w_{e}$ is determined by eq. (\ref{ssq-1}),
\bdm
V(w_{e}) \, = \, R_{e}^{2} \, \left( 1 - \Lambda \, R_{e}^{2} \right) 
\, ,
\edm
and $w''_{e}$, $R^{(4)}_{e}$ and $Q''_{e}$ are obtained from
eqs. (\ref{ssq-0})-(\ref{ssq-3}),
\bdm
(w''_{e})^{2} \, = \, \frac{Q_{e}}{4R^{2}_{e}} V,_{w}(w_{e}) ,
\, \, \, \, \, \, 
R^{(4)}_{e} \, = \, -\frac{4}{R_{e}} (w''_{e})^{2} ,
\, \, \, \, \, \, 
Q''_{e} \, = \, \frac{2}{R_{e}^{4}}(2 V(w_{e}) - R_{e}^{2}) \, . 
\edm

In both cases, the free parameters are the position of the horizon,
$\rho_{e}$, the cosmological constant, the shooting parameter
at the pole, $b$, and two independent shooting parameters at the 
equator (for instance $R_{e}$ and $Q_{e}$).
The values of these quantities are presented in Table.1
for the first five compact solutions.
The shape of the metric functions and the
Yang--Mills amplitude is given in Fig.7 for the $n=3$
solution.

\begin{figure}
\epsfxsize=10cm
\centerline{\epsffile{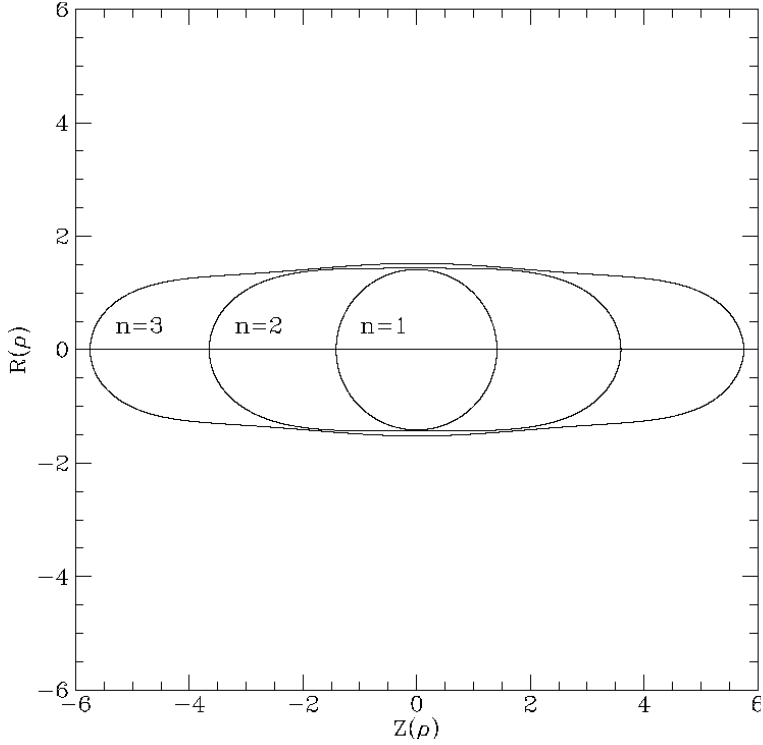}}
\caption{The embedding diagrams for the $n=1,2,3$ compact solutions.}
\label{Fig.8}
\end{figure}

\begin{center}
Tab. 1. Parameters for compact solutions.
\vglue 0.4cm
\begin{tabular}{|l|l|l|l|l|l|l|l} \hline
$n$ & $\Lambda$  & $b$      & $\sqrt{2}\rho_{e}/\pi$    & 
$R_{e}/\sqrt{2}$ & $w_{e} $    & $Q_{e}$ 
\\ \hline
 1  &  0.75      & 0.25     & 1 & 1 & 0 & 1\\
 2  &  0.364244  & 0.429599 & 4.824 & 1.0150 & $-0.5320$ &16.656\\
 3  &  0.293218  & 0.508831 & 15.63 & 1.0757 & 0 & 88.390\\
 4  &  0.270328  & 0.540489 & 39.64 & 1.0483 & 0.23549 & 417.12\\
 5  &  0.260895  & 0.554021 & 88.43 & 1.0485         &    0 &1409.7     \\
 ...&  ...       & ...      & ...      & ...    & ...  & ...  \\
$\infty$&0.25    & 0.569032 & $\infty$ & $1$  & --&$\infty$\\ \hline
\end{tabular}
\end{center}

\vspace{5 mm}

The geometry of the compact solutions may be illustrated 
with the help of embedding diagrams. 
Consider the $3$-dimensional Euclidean space
in cylindrical coordinates $R$, $Z$ and
$\varphi$. 
A surface $S$ of revolution in this space
is characterized by a mapping $\rho \rightarrow
(R(\rho),Z(\rho))$, and the induced metric
on $S$ is
\be
\gtens_{S} \, = \, \left(
R'^{2} + Z'^{2} \right) \, d\rho^2 \, + \, 
R(\rho)^{2} \, d \varphi^2 \, .
\label{5:2}
\ee
On the other hand, the metric of a
spacelike section $S'$ (with $t=t_{0}$ and $\vartheta=\pi/2$)
through the geometry of the compact solutions is given by
\be
\gtens_{S'} \, = \, \frac{1}{Q(\rho)} \, d \rho^{2}
\,+ \, R(\rho)^{2} \, d \varphi^2 \, .
\label{5:1}
\ee
Hence, the two geometries coincide, provided that we choose 
the function $Z(\rho)$ according to
\be
Z(\rho) \, = \, Z(0) \, + \, 
\int_{0}^{\rho}
\sqrt{1-Q(\bar{\rho}) R'^{2}(\bar{\rho})} 
\, \frac{d\bar{\rho}}{\sqrt{Q(\bar{\rho})}} \, ,
\label{5:3}
\ee
where $\rho \in [0,2 \rho_{e}]$.

The embedding diagrams $(R(\rho),Z(\rho))$ 
for the $n=1$, the $n=2$ and the $n=3$ 
compact regular solutions 
are presented in Fig.8. For $n=1$, we obtain the
circle $(R(\rho),Z(\rho)) = 
\sqrt{\kappa} (\sin(\rho / \sqrt{\kappa}),
\cos(\rho / \sqrt{\kappa}))$,
reflecting the fact 
that the manifold in this case is precisely $I \! \! R \times S^{3}$. 
The spatial sections of the solutions with higher values of $n$
resemble prolate ellipsoids (or ``cigars"). 

It is also instructive to draw the embedding diagrams 
for the solutions with horizon. 
In this case, the domain of integration in eq. (\ref{5:3}) 
is $\rho\in [0,\rho_{h}]$, which yields half of the diagram.
At the horizon one has $Q=0$ and therefore $dR/dZ=0$.
Since $R(\rho_{h}) \neq 0$, the horizon corresponds to
the ``throat'' of the geometry, which connects 
the two identical patches of the manifold (see
the conformal diagram in Fig.3).  
The resulting diagrams for several $n=1$ solutions
are presented in Fig.9.
The diagrams show that the throat becomes 
narrower as $\Lambda$ 
tends to the critical value $\Lhig$,
where the manifold splits into two
separate pieces. 

\begin{figure}
\epsfxsize=10cm
\centerline{\epsffile{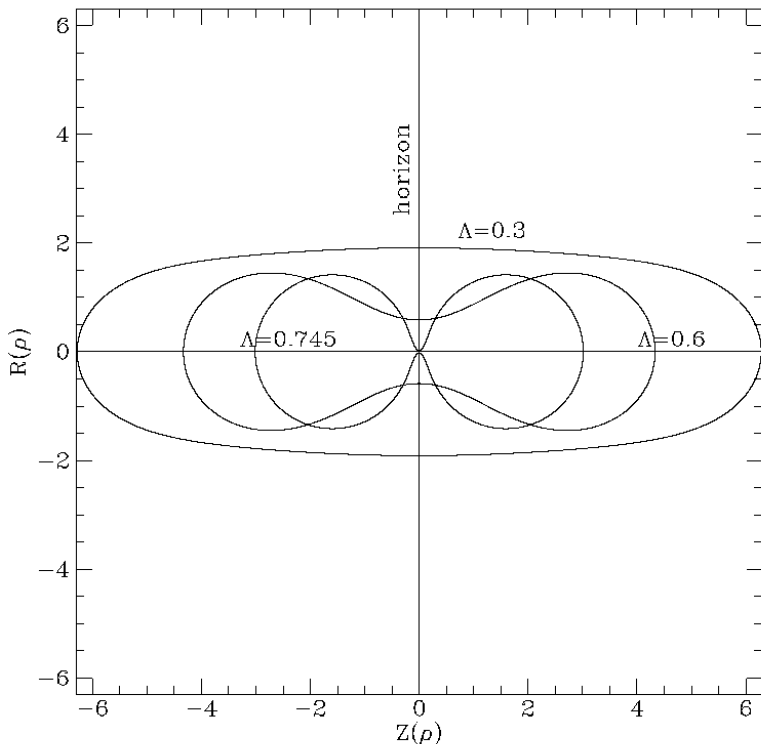}}
\caption{The embedding diagrams for the asymptotically deSitter solution
with $\Lambda=0.3$, $n=1$, and for the $n=1$ bag of gold solutions with
$\Lambda=0.6$ and $\Lambda=0.745$.}
\label{Fig.9}
\end{figure}

\section{Concluding Remarks}

The features of the static, spherically symmetric solutions to
the EYM equations depend critically on the value of the
cosmological constant $\Lambda$. For every node number $n$,
there exists a globally regular, compact solution with
$\Lambda = \Lhig(n)$. For $\Llow(n) < \Lambda < \Lhig(n)$,
the configurations have ``finite size'' and exhibit a horizon.
Finally, for sufficiently small values of the cosmological
constant, $\Lambda < \Llow(n)$, the solutions
generalize the BK solitons surrounded by a  
cosmological horizon.

In this paper, we have restricted our attention
to solutions with a regular center. The 
extension to configurations with an event horizon
is expected to be straightforward. In fact, we have
no reasons to doubt that one will find
a similar classification for these black hole
solutions.

No globally regular solutions seem to exist 
for $\Lambda > \Lhig(n)$. In this case, the metric
function $Q(\rho)$ is everywhere positive and 
diverges as $\rho\rightarrow\rho_{sing}$, where 
$\rho_{sing}$ is the position of the second zero of $R(\rho)$. 
Such solutions may therefore be considered as 
bag of gold configurations without horizon.
When $\Lambda$ is small and negative ($ | \Lambda | \ll 1$), 
the solutions resemble again the BK solitons, however,
they approach the anti--deSitter geometry 
in the asymptotic region.

We have also investigated the stability properties
of the solutions presented in this paper.
The stability analysis for 
these -- asymptotically not flat -- solutions is, 
however, rather involved. In particular,
the fact that the size $R$ of the $2$--spheres develops
a local maximum gives rise to the following difficulty:
Either the pulsation equations assume the form of a regular,
formally self-adjoint system with {\em unphysical} degrees
of freedom, or one isolates the unphysical modes and obtains a 
{\em singular} pulsation equation. 
The methods by which these problems can be solved are
presented in an accompanying paper \cite{stable}, and here we merely
mention the result: all of the solutions described above are unstable. 




\begin{thebibliography}{99}

\newcommand{\AJM}{{\em Am.\ J.\ Math.\ }}
\newcommand{\AM} {{\em Ann.\ Math.\ }}
\newcommand{\AMC}{{\em Ann.\ Math.\ Soc.\ Coll.\ (4th ed.)\ }}
\newcommand{\ANY}{{\em Ann.\ N.Y.\ Acad.\ Sci.\ }} 
\newcommand{\APN}{{\em Ann.\ Phys. \ (NY)\ }}
\newcommand{\AP} {{\em Ann.\ Phys.\ }}
\newcommand{\APK}{{\em Ann.\ Physik.\ }}
\newcommand{\APL}{{\em Ann.\ Physik.\ (Leipzig,\ 4.f.)\ }}
\newcommand{\ASI}{{\em American\ Scientist\ }}
\newcommand{\AIA}{{\em Ann.\ Inst.\ H.\ Poincar\'e\ A\ }}
\newcommand{\AJ} {{\em Astrophys.\ J.\ }}
\newcommand{\BLM}{{\em Bull.\ London\ Math.\ Soc.\ }}
\newcommand{\CQG}{{\em Class.\ Quantum\ Grav.\ }}
\newcommand{\CMP}{{\em Commun.\ Math.\ Phys.\ }}
\newcommand{\CMPshort}{{\em Comm.\ Math.\ Phys.\ }}
\newcommand{\CPA}{{\em Commun.\ Pure \ Appl. \ Math.\ }}
\newcommand{\GRG}{{\em Gen.\ Rel.\ Grav.\ }}
\newcommand{\HPA}{{\em Helv.\ Phys.\ Acta\ }}
\newcommand{\JET}{{\em JETP\ Lett.\ }}
\newcommand{\JDG}{{\em J.\ Diff.\ Geom.\ }}
\newcommand{\JMP}{{\em J.\ Math.\ Phys.\ }}
\newcommand{\JPM}{{\em J.\ Phys.\ A:\ Math.\ Gen.\ }}
\newcommand{\LMP}{{\em Lett.\ Math.\ Phys.\ }}
\newcommand{\MRA}{{\em Mon.\ Not.\ R.\ Astron.\ Soc.\ }}
\newcommand{\NAT}{{\em Nature\ }}
\newcommand{\NPS}{{\em Nature\ (Phys.\ Sci.)\ }}
\newcommand{\NC} {{\em Nuovo\ Cimento\ }}
\newcommand{\NP} {{\em Nucl.\ Phys.\ }}
\newcommand{\NPB}{{\em Nucl.\ Phys.\ B\ }}
\newcommand{\PL} {{\em Phys.\ Lett.\ }}
\newcommand{\PLA}{{\em Phys.\ Lett.\ A\ }}
\newcommand{\PLB}{{\em Phys.\ Lett.\ B\ }}
\newcommand{\PRL}{{\em Phys.\ Rev.\ Lett.\ }}
\newcommand{\PR} {{\em Phys.\ Rev.\ }}
\newcommand{\PRB}{{\em Phys.\ Rev.\ (Sect.\ B)\ }}
\newcommand{\PRP}{{\em Phys.\ Rep.\ }}
\newcommand{\PRD}{{\em Phys.\ Rev.\ D\ }}
\newcommand{\PAW}{{\em Preuss.\ Akad.\ Wiss.\ Berlin,\ Sitz.ber.\ II\ }}
\newcommand{\PTR}{{\em Phil.\ Trans.\ R.\ Soc.\ (London)\ }}
\newcommand{\SDA}{{\em Sitz.ber.\ Deut.\ Akad.\ Wiss.\ Berlin,\ Kl.\ 
Math.-Phys.\ Tech.\ }} 
\newcommand{\SLA}{{\em Proc.\ R.\ Soc.\ London\ Ser.\ A\ }}
\newcommand{\PNA}{{\em Proc.\ Natl.\ Acad.\ Sci.\ }}
\newcommand{\PKN}{{\em Proc.\ Kon.\ Ned.\ Akad.\ Wet.\ }}
\newcommand{\PMS}{{\em Phil.\ Mag.\ (Ser 5)\ }}
\newcommand{\PIA}{{\em Proc.\ Roy.\ Irish\ Acad.\ }}
\newcommand{\PTP}{{\em Progr.\ Theor.\ Phys.\ (Kyoto)\ }}
\newcommand{\RPP}{{\em Rep.\ Prog.\  Phys.\ }}
\newcommand{\RNC}{{\em Riv.\ Nuovo\ Cimento\ }} 
\newcommand{\SJN}{{\em Sov.\ J.\ Nucl.\ Phys.\ }}
\newcommand{\TSM}{{\em Trans.\ Am.\ Math.\ Soc.\ }}


\bibitem{BMK} R.Bartnik,  J.McKinnon, 
\PRL {\bf 61}, 141 (1988). 

\bibitem{BHS96}
O.Brodbeck, M.Heusler, N.Straumann, \PRD {\bf 53}, 754 (1996). 

\bibitem{mavr} N.E. Mavromatos, E.Winstanley, \PRD {\bf 53}, 3190 (1996). 

\bibitem{Hosotani} 
J.Cervero, L.Jacobs, \PLB {\bf 78}, 427 (1978);
M.Henneaux, \JMP {\bf 23}, 830 (1982);
Y.Hosotani, \PLB {\bf 147}, 44 (1984). 

\bibitem{SY1} J.A.Smoller, A.G.Wasserman, S.T.Yau, and J.B.McLeod,
\CMP  {\bf 143}, 115 (1991);
J.A.Smoller, A.G.Wasserman,
\CMP {\bf 151}, 303 (1993);
ibid. {\bf 154}, 377 (1993). 
\bibitem{SY3} P.Breitenlohner, P.Forgacs, D.Maison,
\CMP {\bf 163}, 141 (1994). 


\bibitem{maeda} T.Torii, K.Maeda, T.Tachizawa, \PRD {\bf 52},
R4272 (1995). 

\bibitem{stable} O.Brodbeck, M.Heusler, G.Lavrelashvili,
N.Straumann and M.S.Volkov, 
{\it Stability Analysis of New Solutions of the EYM System with 
Cosmological Constant}, ZU-TH 14/96.



\bibitem{JS81} G.Johnstone Stoney, \PMS {\bf 11}, 381 (1881). 

\bibitem{nariai} H.Nariai, {\it Science Reports of the Tohoku Univ.}
{\bf 35}, 62 (1951). 



\end{thebibliography}
\end{document}